\documentclass[]{pasj02} 
\usepackage[switch,mathlines]{lineno} 
\usepackage{url}
\usepackage{mathcomp}

\usepackage{lmodern}
\usepackage{natbib}

\jyear{2024}
\Received{}
\Accepted{}

\begin{document} 

\title{ Strong ionospheric activity at the MWA site associated with plasma bubble measured by GNSS }

\author{
Shintaro \textsc{Yoshiura}\altaffilmark{1}\email{shintaro.yoshiura@nao.ac.jp},
Yuichi \textsc{Otsuka}\altaffilmark{2}, 
Cathryn M. \textsc{Trott}\altaffilmark{3,4}, 
Dev \textsc{Null}\altaffilmark{3,4}. 
Nozomu \textsc{Nishitani}\altaffilmark{2},  
Keitaro \textsc{Takahashi}\altaffilmark{5}, 
Michi \textsc{Nishioka}\altaffilmark{6}, 
Septi \textsc{Perwitasari}\altaffilmark{6}, 
and Atsuki \textsc{Shinbori}\altaffilmark{2}}

\altaffiltext{1}{Mizusawa VLBI Observatory, National Astronomical Observatory of Japan, 2-21-1 Osawa, Mitaka, Tokyo 181-8588, Japan}
\altaffiltext{2}{Institute for Space-Earth Environmental Research (ISEE), Nagoya University, Nagoya, Japan}
\altaffiltext{3}{International Centre for Radio Astronomy Research, Curtin University, Bentley, WA 6102, Australia}
\altaffiltext{4}{ARC Centre of Excellence for All Sky Astrophysics in 3 Dimensions (ASTRO 3D), Bentley, Australia}
\altaffiltext{5}{Kumamoto University, International Research Organization for Advanced Science and Technology, 2-39-1 Kurokami, Chuo-ku, Kumamoto 860-8555, Japan}
\altaffiltext{6}{National Institute of Information and Communications Technology (NICT), Tokyo, Japan
}


\KeyWords{atmospheric effects, techniques: interferometric, Earth, plasmas, }  

\maketitle

\begin{abstract}
The Earth's ionosphere refracts radio signals, shifting the apparent position of radio sources. Wide-field measurements with a radio interferometer can measure the ionospheric distortion. The Murchison Widefield Array (MWA) has the ability to capture ionospheric structures that are smaller than 100 km in extent. We report unusually strong ionospheric activity in MWA data during a magnetic storm on {2023 December 1}. The duct-like structure (roughly 50 km $\times$ $>$100 km) passes through the MWA field-of-view (FOV) with a velocity of $\sim$ 100 m/s. The offsets of the apparent position of the radio source are more than 1 degree in the MWA observation data at around 180 MHz. By comparing the Total Electron Content (TEC) data obtained from the GNSS receiver network, we have found that the TEC fluctuations represented by a high Rate of TEC change index (ROTI) coincided with the strong ionospheric activity observed by the MWA. This result suggests that unusual ionospheric signatures detected by the MWA could be caused by plasma bubbles extending across Western Australia during a magnetic storm.
\end{abstract}


\section{Introduction}\label{sec:1}

The Earth's ionosphere, which contains ionized media induced by extreme ultraviolet and X-ray radiation from the Sun, is a known issue for low frequency radio interferometric observations \citep[{Ch.~14 in}][]{2017isra.book.....T}. A radio interferometer, consisting of multiple antennas, measures the complex visibility which describes the Fourier transform of the sky brightness of radio sources. The phase difference, caused by the ionosphere, in the visibilities shifts the apparent position of radio sources. Such ionospheric effects scale as {the inverse square of radio frequency}, $\nu^{-2}$, and change on short time scales (a few seconds to minutes). As a result, the quality of imaging of radio sources is limited due to the fluctuations. Conversely, by measuring the position offsets, the radio interferometer can be available to monitor ionospheric activity.

The MWA is a low-frequency radio interferometer \citep{2013PASA...30....7T,2018PASA...35...33W}. The MWA has been used for studying the ionosphere. Previous works used the refractive position offsets of several radio sources in the MWA's wide field of view for revealing spatial gradients in the Total Electron Content (TEC) \citep{Loi3,Loi1,Loi4,2020RaSc...5507106H,2022JATIS...8a1012R}. The study shows a duct-like structure aligned along the Earth's magnetic field and the formation of the field-aligned ducts after the passage of the travelling ionospheric disturbance (TID). The small-scale ionospheric structure is measured via ionospheric scintillation \citep{2022PASA...39...36W}. The classification of ionospheric activity is proposed by \cite{2017MNRAS.471.3974J}. The classification is used as a guide to assess data quality available for radio astronomy. The TEC is {given as} an integration of the electron density {along the pass from the ground to the radio source located far beyond the Earth's ionosphere \citep[{Ch.~14 in}][]{2017isra.book.....T}.} {The global TEC value is obtained using the GPS satellite and the receiver in} Global Navigation Satellite Systems (GNSS) \citep{2002EP&S...54...63O}. The GNSS-based TEC maps can be available for calibration of radio interferometer \citep{2015PASA...32...29A,2016PASA...33...31A}. 

Plasma bubbles, which are a localized electron density depletion in the nighttime equatorial ionosphere, are frequently generated post-sunset through the Rayleigh-Taylor instability \citep[e.g.][]{2009eipp.book.....K}. A plasma bubble has a structure elongating along the geomagnetic field lines \citep[e.g.][]{2002GeoRL..29.1753O}. To generate a plasma bubble, eastward electric fields play an important role. At the evening terminator, an intense electric field, so-called pre-reversal enhancement (PRE), is generated through the E and F region coupling process. During the main phase of a geomagnetic storm, the convection electric field at high latitudes likely penetrates into the equatorial ionosphere \citep[e.g.][]{2008JGRA..113.6214K} . The penetration electric field is eastward in dayside and duskside, and westward in the nightside. At the evening terminator, the penetration electric field superimposes on the PRE to induce a large eastward electric field, resulting in favourable conditions for plasma bubble generation. For this condition, plasma bubbles likely extend to mid-latitudes \citep[e.g.][]{2006GeoRL..3321103M,2016GeoRL..4311137C,2021JGRA..12629010S}. In the plasma bubbles, the plasma density irregularities with various scale size coexist {\citep{1978JGR....83.4219B}}. Consequently, plasma bubbles are observed as fluctuations of TEC obtained from the GNSS data \citep[e.g.][]{2008JGRA..113.5301N}.

On {2023 December 1}, during a magnetic storm {\citep{2024GeoRL..5108778K,2024JGRA..12932430S}}, the MWA telescope, located in Western Australia (WA), observed one of the most significant ionospheric phase errors reported in its past observations. During this event, plasma bubbles were observed over the MWA as TEC fluctuations observed by the GNSS receivers.

This paper proceeds as follows. In Section \ref{sec:2} we explain the data observed with the GNSS and MWA. In Section \ref{sec:3} we show the results of {Rate Of TEC change Index (ROTI)} with the distribution of position offsets of radio sources and clear ionospheric distortion in the radio images. The summary is given in Section \ref{sec:4}.

\section{Observations and Analysis}\label{sec:2}

Data were obtained from more than 8,000 GNSS receivers around the world on {2023 December 1}. The list of the data providers is described at \url{https://stdb2.isee.nagoya-u.ac.jp/GPS/GPS-TEC/gnss_provider_list.html} {and in \cite{2020JGRA..12526873S}}. TEC for each pair of satellite and receiver is calculated from the carrier phase and pseudorange data for the dual-frequencies of multi GNSS. ROTI is widely used to detect plasma density irregularities in the F region \citep[e.g.][]{1997GeoRL..24.2283P, 2008JGRA..113.5301N}. ROTI is defined as a standard deviation of ROT (Rate Of TEC change) in 5 minutes. ROT is the differential of TEC at 30-second intervals, and represents the magnitude of TEC change in each minute.  The obtained ROTI is converted to a vertical ROTI by multiplication of a slant factor, defined as $\tau_0$/$\tau_1$, where $\tau_1$ is the length of the ray path between 250 and 450 km altitudes and $\tau_0$ is the thickness of the ionosphere (200 km) for the zenith path. In order to distinguish temporal and spatial variations of TEC, the ROTI are mapped onto an ionospheric {layer} at a 300 km altitude in the geographical coordinates with a horizontal cell of 0.25$^\circ$ $\times$ 0.25$^\circ$ in latitude and longitude. The ROTI indicates existence of the ionospheric irregularities with scale size from 3 to 30 km \citep[e.g.][]{2008JGRA..113.5301N}.

The {MWA} is a low-frequency radio interferometer built for various radio astronomical science \citep{2019PASA...36...50B}. The MWA is located at Inyarrimanha Ilgari Bundara, the CSIRO Murchison Radio-astronomy Observatory (MRO) in {WA}. The MWA was in operation on {2023 December 1}, obtaining data for measuring the HI signal from the Epoch of Reionization (EoR). On that day, 145 out of 256 tiles, each consisting of 16 antennas, were used in a compact array configuration with a maximum baseline of {992~m}. The frequency bandwidth was 32 MHz, centred at approximately 182 MHz.

The observations have covered two target fields; the so-called EoR1 (RA=4h, Dec=-27$\tcdegree$) and EoR3 (RA=1h, Dec=-27$\tcdegree$) fields. The antenna gain solutions are derived with the Hyperdrive (\url{https://github.com/MWATelescope/mwa_hyperdrive}) calibration pipeline. {We run the hyperdrive calibration for each 112~s data chunk using the sky model of the target field (EoR1 and EoR3). We found a wonky frequency structure in the gain amplitude and phase after 12:50 (UT). This indicates that the observed sky is modulated and sky-based calibration cannot derive proper gain values \citep[][]{Barry2016,2017MNRAS.470.1849E,2019ApJ...875...70B}. Thus, we evaluate an averaged gain solution between the gain solutions from 11:56:46 (UT) to 12:28:46 (UT) where the gain solution shows stable frequency structure. The averaged gain solution is applied to all of the data to create the calibrated data.} The calibrated data are converted to images of {2048 $\times$ 2048 pixels, with a pixel resolution of 1 arcmin after WSCLEAN (Offringa et al. 2014) with a uniform weighting and the maximum iterations of 10000.} The calibrated data are also processed with the peeling mode of the Hyperdrive, which measures the offset of the position of bright sources against the catalogue position. The ionospheric offset{, which is the difference of apparent position of radio source from the cataloged position,} is measured every {16~s} by following the method described in \cite{2008ISTSP...2..707M} by assuming all MWA tiles measure a single ionospheric layer with gradients toward the source direction\footnote{For conversion from the hyperdrive outputs to the positional offsets, we use the method given in \cite{2017MNRAS.471.3974J} and \url{https://gitlab.com/chjordan/cthulhu/}.}.  We run Hyperdrive for measuring the offset of 4000 sources after the subtraction of 8000 objects.

\section{Results {and} Discussion}\label{sec:3}

\subsection{Arrival of plasma bubble in ROTI map and {ionospheric offset}}

The magnetic storm {occurred on 2023 December 1 was} induced by the coronal mass ejection (CME) associated with the M9.8 solar flare that occurred on {2023 November 28} \citep{2024GeoRL..5108778K}. The Dst index reached {the} minimum of -108 nT at 14 UT on {2023 December 1}. The plasma bubble occurred above the equator due to a Rayleigh-Taylor instability. {\cite{2024JGRA..12932430S} reported the extension of the plasma bubble at high latitude in the northern hemisphere on that date using data measured from GNSS data, ionosondes, and radars.}

\begin{figure}
\centering
\includegraphics[width=8cm]{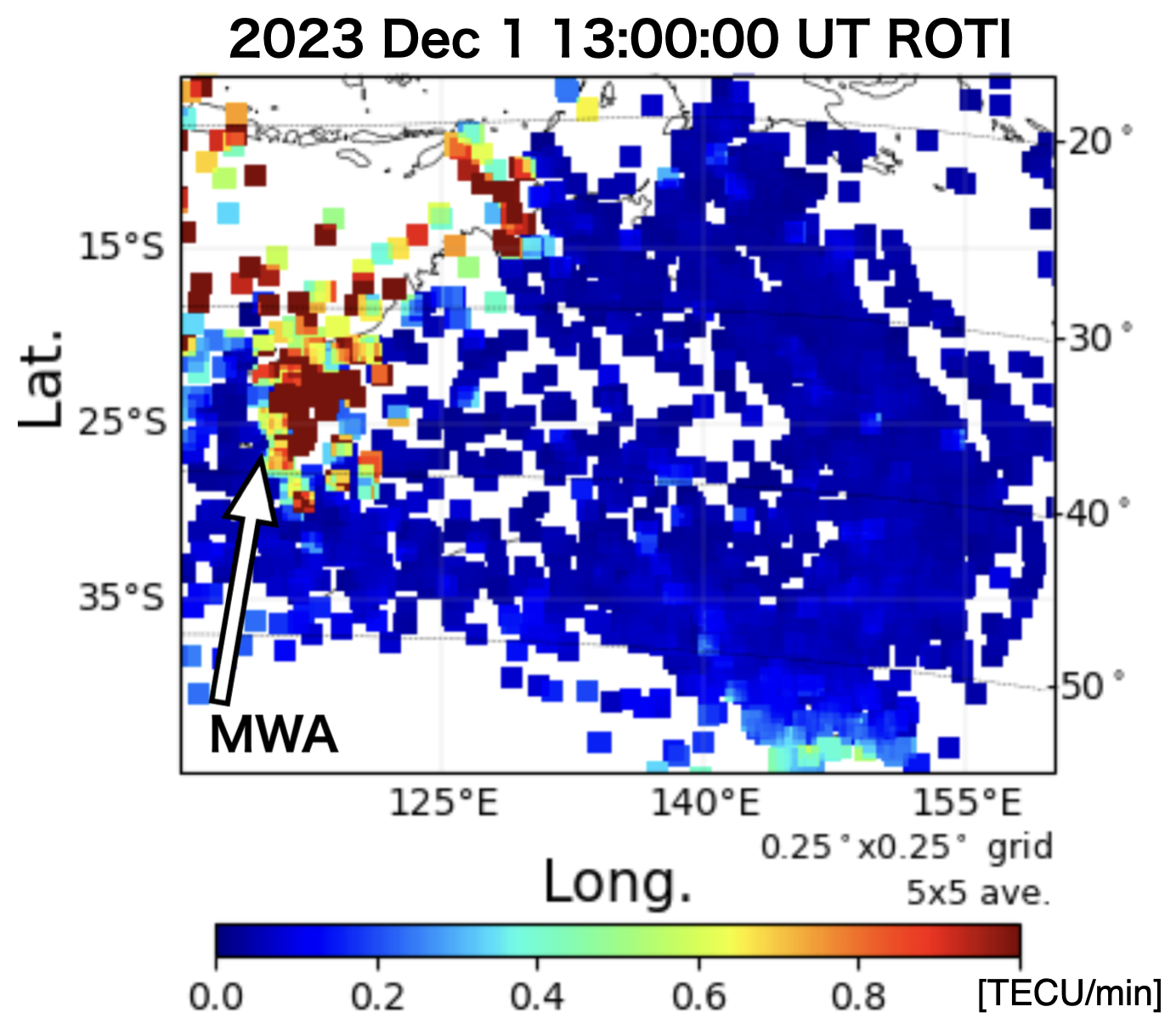}
\caption{The ROTI map of the Oceania region at 13:00 (UT) on 1st Dec 2023 {when the large cluster of high ROTI values arrived over the MWA site. The grid sizes of ROTI map are 0.25x0.25 degrees and averaged values using 5x5 grids are shown.} The white arrow roughly indicates the MWA site at (lat, long) = (-26.70, 116.67). {Alt text: The colour contour of the ROTI in units of TECU/min. The x-axis shows longitude from approximately 110 degrees to 160 degrees. The y-axis shows southern latitude from 10 degrees to 45 degrees.}}
\label{fig:roti-l}
\end{figure}

Figure~\ref{fig:roti-l} shows that the high ROTI region exists over the MWA site. The movement of the high ROTI region can be seen in supplemental material S1. The large scale patch of high ROTI propagates to WA at around {13:00 (UT)}. 

\begin{figure*}
\centering
\includegraphics[width=15cm]{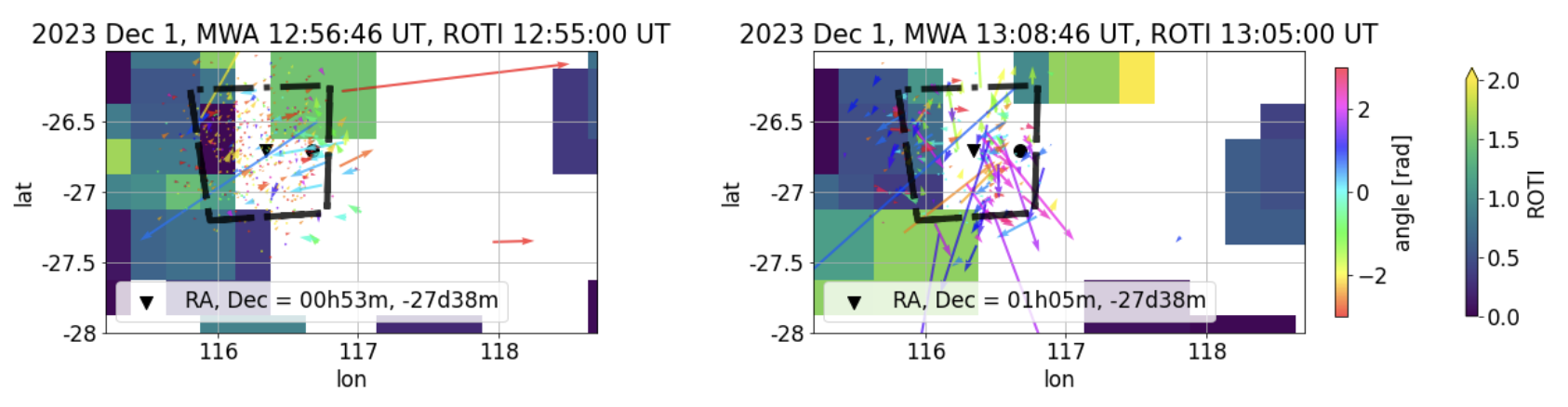}
\caption{{The ROTI map} from GNSS observation on 1st Dec 2023 above WA. {Colour of tiles indicates the strength of ROTI in TECU/min as indicated by the color bar in the right.} The {positional offsets of radio sources from the cataloged position} are over-plotted on the ROTI map. {Assuming an ionospheric layer at an altitude of 300~km, the source position in Geographic coordinates is evaluated as the vertical projection of the intersection point between the ionosphere and the line of sight to the source.} The vectors start from the direction of the radio sources. The colour of the vector indicates the direction; the length of the vector is scaled to 30 times larger than actual size for clear visualization. {Left panel shows the ROTI and vectors at the beginning of significant ionospheric distortions. Right panel corresponds to the time when a plasma bubble crossing the primary beam of the MWA.} {For reference, we mark the pointing centre of the MWA as a black inverted-triangle. The 20 $\rm deg$ $\times$ 20 $\rm deg$ region, which roughly corresponding to the FoV of MWA, around the pointing centre is also illustrated as black dashed line.} {Alt text: The x-axis shows longitude from approximately 115 degrees to 118.5 degrees. The y-axis shows latitude from -28 degrees to -25 degrees. }}
\label{fig:roti}
\end{figure*}

Figure~\ref{fig:roti} shows the zoomed ROTI map around the MWA site measured with GNSS. The vector fields show the offsets of radio sources observed by the MWA. The movie of the images is given as supplemental material S2. From 12:50 to 13:30 (UT), we found significant changes in the vectors describing the offset of radio sources from their catalogue position, within the EoR3 field. At 12:56 (UT), in the left panel of figure~\ref{fig:roti}, the Hyperdrive calibration pipeline is not able to measure the offset of radio sources at the eastern region. This is because the offset is larger than a threshold value. The typical standard value of the threshold is a few arcmin, but we used $\sim$ 1 degree for approximately capturing the large ionospheric offsets. Note that the method used in Hyperdrive does not work properly if the ionospheric deviations are larger than a few arcmin. At 13:08 (UT), in the right panel of figure~\ref{fig:roti}, the large-scale patch of high ROTI comes from the northeast. The vector field shows that there are large vectors and a vertical region where the vectors are sparse. The median value of radio source position offset is 1 arcmin. The figure~\ref{fig:roti-l} and figure~\ref{fig:roti} indicate that the plasma bubble extended to WA. 

Before the arrival of the plasma bubble, at 12:00 (UT), the ROTI is low. The median value of the offsets is roughly 0.1 arcmin at $\approx$182MHz. This value is a typical value for the calm state of the ionosphere \citep{2017MNRAS.471.3974J}. At 12:30 (UT), the strong ROTI appeared north of the MWA. The ROTI map might indicate that the plasma bubble has arrived over WA and the patch of higher ROTI moves toward the north. However the ionospheric offsets do not show any reaction. The reason for this calmness of radio data is unknown. It might be due to some systematic error of GNSS data at a low elevation angle.

After the plasma bubble passes through, from 13:30 UT, the MWA starts to observe the EoR1 field (RA, Dec=4h, -30 degrees). Although the large-scale patch of high (larger than 1) ROTI has already passed over the MWA site, we still see large offsets and the median value of offset is roughly 1 arcmin. After 15:00, the offsets start to show straight and parallel structures. The field-aligned structure has been reported in previous works \citep{Loi1}. The origin of the structure could be the density fluctuations aligned with the magnetic field of the Earth. The results might indicate that the small-scale (a few km) structures, observable with the ROTI, in the plasma bubble have calmed. On the other hand, the large-scale (more than 10 km) structures might be caused after the plasma bubble passes through and becomes visible by the MWA.

\subsection{Duct-like large structure in radio images and split source}

\begin{figure*}
\centering
\includegraphics[width=15cm]{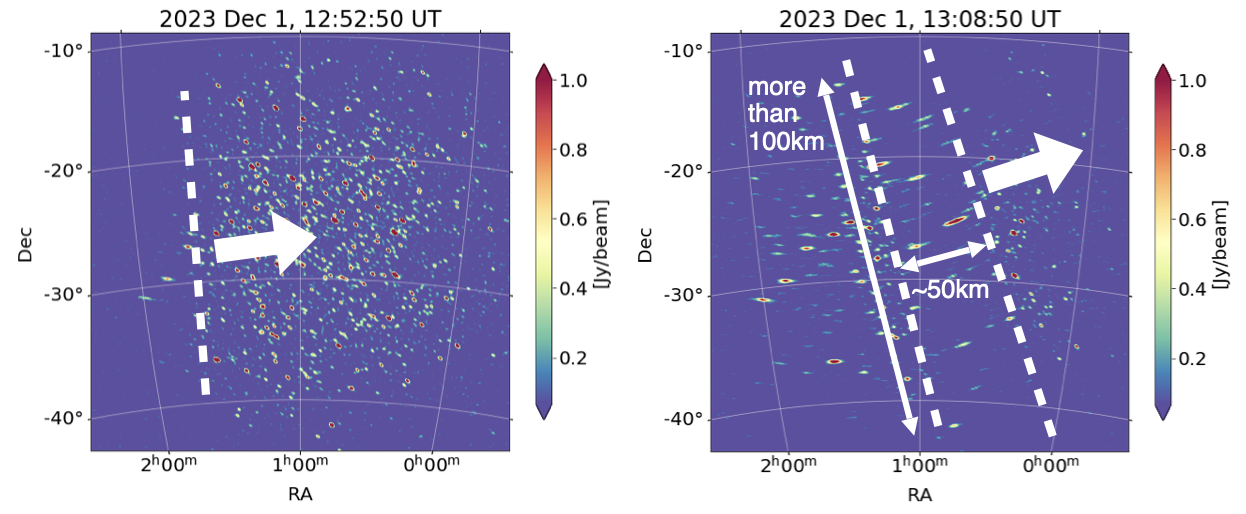}
\caption{Plot of radio sky using 2 minutes of MWA data having applied the averaged calibration solution. These are cleaned images produced using the WSCLEAN software with uniform weighting. For clear visualization of the ionospheric effect, we limit the colour scale from 3 $\sigma_{\rm rms}$ to 1 [Jy/beam]. {The image rms, $\sigma_{\rm rms}$, is 0.022 [Jy/beam] measured at outside of primary beam and at $\approx$ 12:00:00 (UT).} {Note that the MWA has significantly wide field of view (approximately 400 $\rm deg^2$)}. Full animation of the images is available as supplemental material S3. White dashed lines indicate the start and end point of smearing of radio sources. The white vector shows the propagation direction of the duct-like structure, the size of which is derived assuming the altitude of the ionospheric layer is above 300 km. {Alt text: Three coloured images. The x-axis shows RA from approximately -23h30' to 2h30'. The y-axis shows Dec from approximately -42 degrees to -10 degrees. }}
\label{fig:sky}
\end{figure*}

Figure~\ref{fig:sky} shows radio images using 2 minutes of MWA data. At 12:52 (UT), the shape of sources starts to change at RA=2h. The smearing is caused by the ionospheric phase error within 2 minutes. The region with clear smearing moves to the north-west. The right panel in figure~\ref{fig:sky} shows the radio image at 13:08(UT). There is a duct-like region where the smearing is clearly visible. The duct-like region has a significant gradient of TEC over 5 degrees to 10 degrees which might correspond to a plasma bubble with a size of 50 km (east-west), assuming that the height of the ionospheric layer is 300 km. The duct-like structure moves $\sim$ 100 km within 15 minutes. The velocity of 100 m/s does not conflict with the typical value for a plasma bubble. The duct-like structure spans from north-northeast to south-southwest and moves from east to north-west. This feature can be explained if the plasma bubble with the backward C-shape \citep{2009JGRA..11411302K} moves towards the south-west. 

The Supplemental Material S3 is a movie of radio images from the EoR3 observing field from 11:44 to 13:30 (UT). Before the arrival of the plasma bubble, the radio sky is very calm. After the duct-like structure passes through ($\approx$ 13:28 UT), the quality of the image is low. This indicates that the phase error due to the ionosphere is still significant.

A large-scale smeared region can be seen at 15:14 (UT) in the image of the EoR1 field (given as supplemental material S4) and does not show clear propagation. The vector field indicates that the Earth's magnetic field is aligned with the TEC structure, as mentioned. Thus, the large-scale smeared region at the time is caused by the alignment of the field with the ionosphere. On the other hand, we see clear scintillation of radio sources. Thus, the small-scale fluctuations of the ionosphere still somewhat remain, even 4 hours after the arrival of the plasma bubble front. 

\begin{figure*}
\centering
\includegraphics[width=16cm]{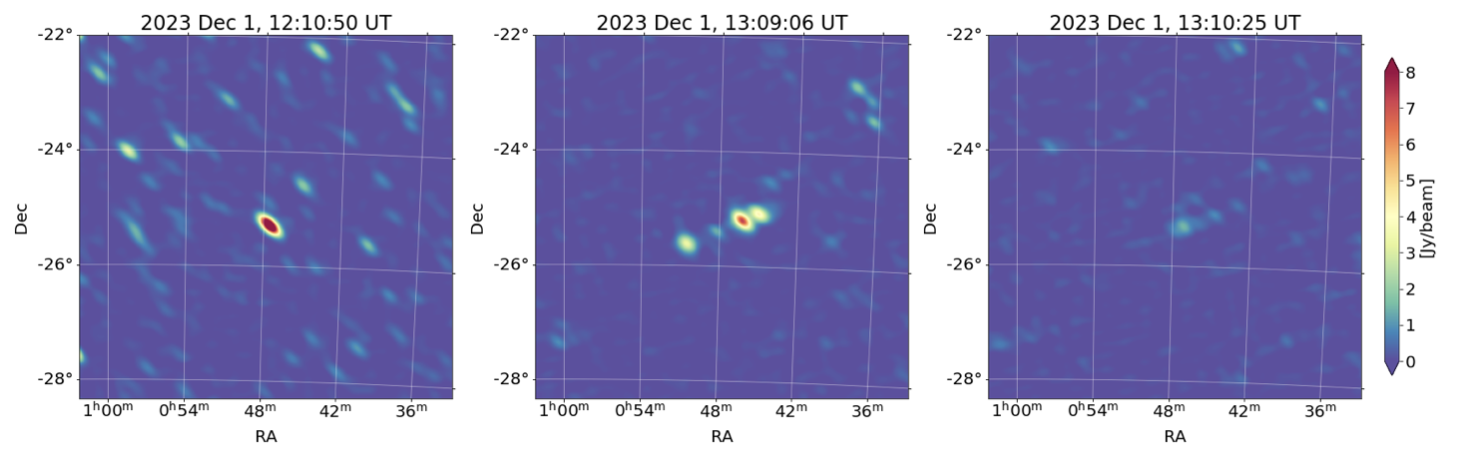}
\caption{Zoomed images of J004733-251710 at (RA, Dec) = (0h47', -25.29). The images are created using the WSCLEAN software with uniform weighting. We limit the colour scale from 0 to 8 [Jy/beam] . The left panel is created using two minutes of data with a calm ionospheric state, for reference. The middle and right panels are created with active ionospheric conditions. The time resolution of these images is 4 seconds. The full animation is given as supplemental material S5. We limit the colour scale from 0 to 1 [Jy/beam] in S5. {Alt text: Three coloured images. The x-axis shows RA from approximately 1h00' to 1h38'. The y-axis shows Dec from approximately -28 degrees to -22 degrees. } }
\label{fig:src}
\end{figure*}

To look at the phase shift in short time steps, we focus on one of the brightest radio sources, J004733-251710, located at (RA, Dec) = (0h47', -25.29). The flux density is 9.35 Jy at 180MHz. The figure~\ref{fig:src} shows 4-second snapshots and an animation is given as supplemental material S5. When the duct-like structure crosses the source direction, the source splits into multiple sources (middle panel of figure~\ref{fig:src}). As a reference, we plot the 2-minute snapshot in the same region using one of the datasets with calm ionospheric activity in the left panel of figure~\ref{fig:src}. There are no sources responsible for the splitting components except J004733-251710. The position of the components varies with the time. These facts show rapid variation of the gradient of the ionospheric layer, {and the number of splitting components and its position are gradually and sometimes suddenly varying within a few seconds.} Under the assumption of a single ionospheric layer, the offset can be described as $\Delta \theta \approx -40.3 / (\nu^2) \nabla_{\perp} {\rm TEC}$ \citep{2016JGRA..121.1569L}. As the apparent offset is roughly 1 degrees, the $\nabla_{\perp} {\rm TEC}$ is 1.5~$\rm[TECU/km]$.

{The plasma bubble contains small scale structures ranging from a few cm to a few tens of km \citep{1978JGR....83.4219B, GL007i010p00848}. The complicated small scale structure of the plasma bubble causes significant refraction and diffraction distorting distort the phase of incoming signal. As the ionosphere structure is uniform along the magnetic field line (see e.g. \cite{Loi1}), the spatial fluctuation align vertically with the geomagnetic field lines. The aligned components from bottom left to top right indicate that the small scale structures of ionized plasma act as a diffraction grating.}

{The time variation of components of the splitting source is caused due to the movement of the plasma bubble. Given a propagation velocity of 100~m/s, the figure indicates the presence of structures ranging from a few hundred meters to a few km. }

{At 13:10(UT), as shown in the right panel of figure~\ref{fig:src}, the apparent source flux decreases severely \footnote{The ionospheric absorption should not be responsible for the decrease. The absorption is only $\approx$ 0.1~dB (roughly 2~\%) at 100MHz with the TEC of $5$ TECU \citep[{Ch.~14}]{2017isra.book.....T}. Furthermore, the plasma bubble has a low density of electrons.}. Flux mitigation of 1 $\sim$ 2 orders of magnitude has been reported in the plasma bubble \citep{2006JMeSJ..84A.343O}. The flux variance has also been reported in the MWA \citep{2022PASA...39...36W}, although they are not related to plasma bubbles. The flux mitigation is likely caused by the ionospheric scintillation. The scintillation is sensitive to structures on the Fresnel scale. At 180~MHz and assuming the ionosphere height of 300~km, the fresnel scale is roughly 1~km. The fresnel scale is consistent with the size of the structure indicated by the time variation of components.
}

\section{Summary}\label{sec:4}

This work reports high ionospheric activity in the GNSS-based ROTI map over Western Australia, and the associated significant distortion in the radio data observed with the MWA. The radio observations and ROTI maps indicate the arrival of a plasma bubble. The MWA's large field of view captured the propagation of a duct-like structure which is 50 km wide to the east-west and more than 100 km to the north-south. The positional shift of 1 degree is one of the most significant ionospheric phase errors reported in the past MWA observations. The detailed interpretation of the phenomenon in the GNSS and MWA data will be discussed in our future work.

\section*{Supplementary data} 

The following supplementary data is available at PASJ online.

S1. The movie of the ROTI map of the Oceania region on 1st Dec 2023. 

S2. The movie of the ROTI map and the ionospheric offsets. The {map with colored tiles} shows the ROTI and the vectors indicate the offset of radio sources {from the catalogue positions}.   

S3. The movie of the radio sky using 2 minutes of MWA data from 11:42 to 13:30 (UT). For clear visualization of the ionospheric effect, we limit the colour scale from 0.066 to 1 [Jy/beam].

S4. The movie of the radio sky using 2 minutes of MWA data from 13:34 to 17:30 (UT). For clear visualization of the ionospheric effect, we limit the colour scale from 0.066 to 1 [Jy/beam].

S5. The movie of the zoomed image of J004733-251710 using 4 seconds of MWA data. We limit the colour scale from 0 to 1 [Jy/beam].

\begin{ack}
This scientific work uses data obtained from \textit{Inyarrimanha Ilgari Bundara} / the Murchison Radio-astronomy Observatory. We acknowledge the Wajarri Yamaji People as the Traditional Owners and native title holders of the Observatory site.

Global GNSS-TEC data processing has been supported by JSPS KAKENHI Grant Number 16H06286. GNSS RINEX files for the GNSS-TEC processing are provided from many organizations listed by the webpage
(\url{http://stdb2.isee.nagoya-u.ac.jp/GPS/GPS-TEC/gnss_provider_list.html}).
\end{ack}

We acknowledge to MWA EoR team members for their useful comments in the early stage of this work. 

\section*{Funding}
This work is supported by JSPS KAKENHI (Grant Number 21J00416(SY), 24K17098(SY), 24K00625(SY), 22K21345(YO), 21H04518(YO), 20H00197(YO), 20H00180(KT), 21H01130(KT), 21H04467(KT), and 24K07112 (AS)), JSPS Bilateral Joint Research Projects no. JPJSBP120247202, and JSPS476 Core-to-Core Program, B. Asia-Africa Science Platforms. KT is partially supported by JSPS Bilateral Joint Research Projects of JSPS, and the ISM Cooperative Research Program (2023-ISMCRP-2046). This research was partly supported by the Australian Research Council Centre of Excellence for All Sky Astrophysics in 3 Dimensions (ASTRO 3D), through project number CE170100013. The International Centre for Radio Astronomy Research (ICRAR) is a Joint Venture of Curtin University and The University of Western Australia, funded by the Western Australian State government. Establishment of CSIRO’s Murchison Radio-astronomy Observatory is an initiative of the Australian Government, with support from the Government of Western Australia and the Science and Industry Endowment Fund. Support for the operation of the MWA is provided by the Australian Government (NCRIS), under a contract to Curtin University administered by Astronomy Australia Limited. 
\section*{Data availability} 
The MWA data used for calculating the ionospheric activity in the study will be available at MWA ASVO via \url{https://asvo.mwatelescope.org} after restriction period, under the MWA DATA ACCESS POLICY. GNSS RINEX files are available from the data providers listed in
\url{http://stdb2.isee.nagoya-u.ac.jp/GPS/GPS-TEC/gnss_provider_list.html}.






\bibliographystyle{apj}

\bibliography{bibtext}

\end{document}